%% file: main.tex
\documentclass[conference]{IEEEtran}
\IEEEoverridecommandlockouts
\usepackage{preamble}
\usepackage{listings}[float]
\usepackage{balance}

\makeatletter
 \let\old@ps@headings\ps@headings
 \let\old@ps@IEEEtitlepagestyle\ps@IEEEtitlepagestyle
 \def\confheader#1{%
 \def\ps@headings{%
 \old@ps@headings%
 \def\@oddhead{\strut\hfill#1\hfill\strut}%
 \def\@evenhead{\strut\hfill#1\hfill\strut}%
 }%
 \def\ps@IEEEtitlepagestyle{%
 \old@ps@IEEEtitlepagestyle%
 \def\@oddhead{\strut\hfill#1\hfill\strut}%
 \def\@evenhead{\strut\hfill#1\hfill\strut}%
 }%
 \ps@headings%
 }
 \makeatother
\confheader{%
This paper is accepted at the IEEE International Mediterranean Conference on Communications and Networking (MeditCom) 2023.
}

\begin{document}


\title{
Knowledge-based Intent Modeling for Next Generation Cellular Networks}

\author{\IEEEauthorblockN{Kashif Mehmood, Katina Kralevska, and David Palma}
\IEEEauthorblockA{\textit{Department of Information Security and Communication Technology (IIK)} \\
\textit{Norwegian University of Science and Technology - NTNU, Trondheim, Norway}\\
Email: \{kashif.mehmood, katinak, david.palma\}@ntnu.no}
}

\maketitle

\import{./Sections}{0.5-Abstract}
\import{./Sections}{1-Sec1}
\import{./Sections}{2-Sec2}

\import{./Sections}{3-Sec3}
\import{./Sections}{4-Sec4}

\import{./Sections}{5-Sec5}

\vspace{-0.5cm}

\bibliographystyle{ieeetr}
\typeout{}

\balance
\bibliography{refs}

\end{document}

%% file: Sections/0.5-Abstract.tex
\begin{abstract}


Intent-based networking (IBN) facilitates the representation of consumer expectations in a declarative and domain-independent form.
However, mapping intents to service and resource models remains an open challenge.
IBN requires handling existing system data in a structured yet flexible structure way. 
Knowledge graphs provide an efficient conceptual framework for constructing contexts and organizing known information. We utilize knowledge graphs to construct a knowledge-based for modeling of intents in the networking domain. In addition, this work also proposes a knowledge-based intent modeling and processing methodology, extending the standardized intent common model proposed by TM Forum for next-generation cellular networks and services.
The proposed knowledge-based IBN approach is demonstrated for next-generation cellular services, validating its potential.
\end{abstract}

%% file: Sections/1-Sec1.tex
\section{Introduction} \label{sec:intro}
\Gls{ibn} is considered a key enabler for autonomous network and service management of next generation networks~\cite{intentdefsrfc9315,MEHMOOD2023109477}. 
It leads toward networks that are fundamentally simpler to manage and operate, requiring only minimal outside intervention. 
A crucial design goal with \gls{ibn} management is the creation of a contextual model to describe, comprehend, and deploy a high-level intent as a service. 
This process involves mapping functionalities between the domain-specific and agnostic representations of user input for network management decisions. 
The processing of a domain-agnostic context model is performed by a network administrator with closed-loop information exchange with the deployment infrastructure and intent-generating user. The user input is also refined to create dynamic network control policies after deliberation with the available service offerings and other stakeholders. These policies can be translated into domain-specific configurations based on the underlying infrastructure. 

In order to provide interoperability between different domains, the process of intent consumption by the network requires a standardized intent representation model. In 2022, the \gls{3gpp} \cite{3gppintent28312} and \gls{tmf} \cite{tmforum-intent-common-model-tr290} have standardized an \gls{icm} consisting of a set of expectations, each of which is defined in terms of a set of parameters, targets and associated restrictions for the network. 
Intent language models have been proposed with pre-defined vocabulary inspired by network or operational named entities. The choice of a natural language~\cite{intent-lang-nemo, 10.1145/3229584.3229590, lumi} or domain-specific language~\cite{ietf-aut-net-rfc7575} for intent expression remains an interesting and challenging direction with tradeoffs in either case. However, the modeling of intents for utilization by network administrators remains a fundamental challenge. This paper considers the established data organization frameworks --- \gls{owl} and \gls{rdf} for modeling intents as knowledge instances. Previous works already used knowledge graphs for intent-driven networking~\cite{intent-ont-kid,intent-ont-indira}. However, these works lack a standardized intent model due to specific information classes for different knowledge graphs which make them limited to specific domains. 


This work builds on \gls{tmf} and \gls{3gpp} definitions of intent and associated relationships within the intent lifecycle to organize different sources of data in the networking and service orchestration domains. The proposed knowledge organization framework utilizes ontologies for intent, resource, and service models for \gls{mc} and \gls{nmc} services. These models are used to propose an intent processing engine with contextual reasoning of the service requirements for intent translation into specific service intents for deployment via a slicing engine. The contributions of the paper are listed below:

\begin{itemize}
    \item Implementation of an intent ontology and description model by extending the standardized model proposed by \gls{tmf}~\cite{tmforum-intent-common-model-tr290}. This includes the definition of an intent model that is domain agnostic and provides sufficient flexibility for different application use cases. 
    \item Ontology and service description extension models to represent mission-critical and non-mission critical services using standardized performance metrics and \glspl{slo}~\cite{3gpp-5gsystem-23-501}. We make these models publicly available~\cite{intentRDF23}.
    \item A knowledge graph-based IBN framework for processing and translation of intents into domain-specific service requests and deployment of the services.
    \item Demonstrate the mapping of domain-agnostic intents in a network simulator, using a non-standalone \gls{5g} architecture for validating the deployment of intents as services. 
\end{itemize}

The rest of the paper is organized as follows. Section \ref{sec:know-models} presents the proposed knowledge base consisting of ontological descriptions. Section \ref{sec:framework} presents the \gls{ibn} framework for management and orchestration of intents. Section \ref{sec: results} covers the proposed proof-of-concept validation framework for the deployment of intents as requested services. Section \ref{sec: conc} concludes the paper.

%% file: Sections/2-Sec2.tex
\section{Proposed Knowledge Models for Intents and Services} 
 \label{sec:know-models}
An intent has been defined as “\textit{a set of operational goals (that a network should meet) and outcomes (that a network is supposed to deliver), defined in a declarative manner without specifying how to achieve or implement them}”~\cite{intentdefsrfc9315}.
An intent can essentially be treated as a knowledge object with an associated lifecycle. This knowledge must also be provided to the system in a way that automated reasoning processes can convert it into a suitable system action. This means that learning about the consumer's expectations must be communicated, disseminated, and regulated in a standardized manner. The knowledge required to be expressed in intent is described as follows:

\begin{itemize}
    \item \textit{Functional requirements:} consist of the type of service to be delivered along with the expected functionality for the users;
    \item \textit{Non-functional requirements:} consist of the information required to explain expectations of the required service e.g., \glspl{kpi} and associated metrics;
\end{itemize}

The representation of intent knowledge must follow a structured approach, and knowledge graphs~\cite{know-graphs} provide an excellent information organization tool.

\subsection{Knowledge Graphs in \gls{ibn}}
Intents are considered knowledge objects with well-defined expressions based on intrinsic vocabulary \cite{intentdefsrfc9315} and semantic roles for different entities like requirements, objectives, and associated constraints. A key goal in any intent model is the standardization of the intent model to provide interoperability for multi-domain and application use cases. 

\begin{figure}[!tbp]
\begin{center}
\includegraphics[width=0.49\textwidth]{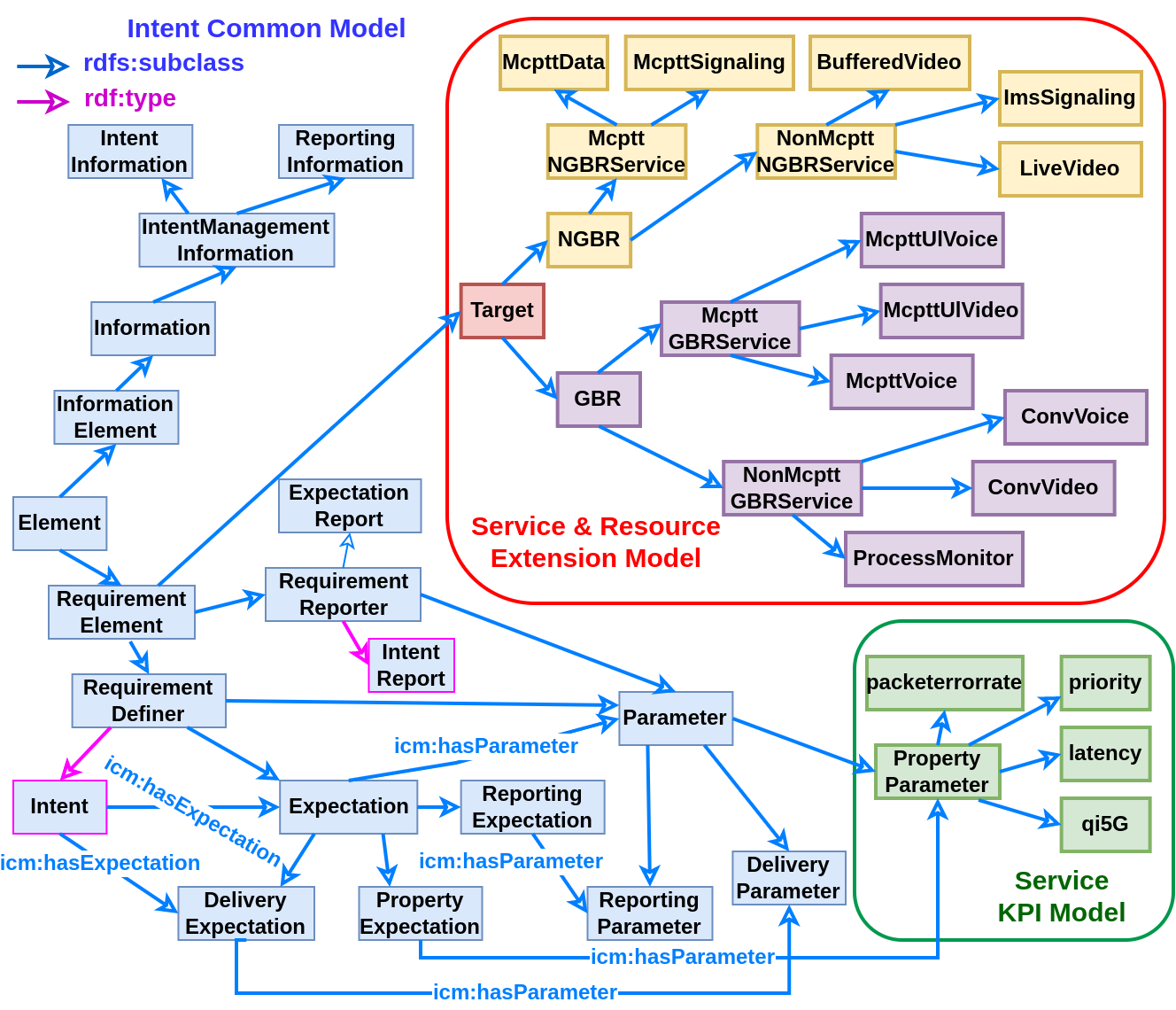}
\end{center}
\caption{Proposed intent model with service, resource and KPI extensions.}
\label{image: tmf-icm-mcptt-nmcptt}
\end{figure}

The standardized ICM \cite{tmforum-intent-common-model-tr290} is a basic intent description template, represented in RDF, that consists of knowledge entities necessary to describe any type of intent ranging from domain-specific to multi-domain high-level intents. We modify the \gls{icm} with additional information and implement it as a knowledge graph depicted in \figurename~\ref{image: tmf-icm-mcptt-nmcptt}. The proposed extensions are explained in the following subsections.

\subsection{Proposed Service and Resource, and Service KPI Extension Models}
A communication service is expected to have a distinct set of associated deliverable metrics and \glspl{kpi} from a variety of standard well-defined performance metrics \cite{3gpp-5gkpis-28-554}. Hence, a service model consists of a set of expectations that need to be fulfilled by the service provider for the service consumers as per \gls{sla}. These sets of expectations are defined as distinct \glspl{slo} mapped onto a set of parameters such as availability, reliability, latency, and bandwidth. 

We utilize \gls{icm} as the basis for our proposed knowledge base consisting of an implementation of intent extension models for the \gls{5g} \gls{mc} and \gls{nmc} services listed in Table~\ref{tab:services}. The proposed extension models ensure compatibility by complementing several information classes including \textit{icm:Expectation, icm:Parameter} and, \textit{icm:Target} from the \gls{icm}. These extended information classes are utilized to incorporate domain-specific data for the services and resources for the intent deployment in the network. The \gls{rdf} classes in the extension models in \figurename~\ref{image: tmf-icm-mcptt-nmcptt} are described as follows:

\subsubsection{Service and Resource Model}
The resource model is extended via the \textit{icm:Target} class. This is the intended target for the deployment of an intent objective (e.g., a service). In the proposed framework, this represents the necessary resource type (i.e., targetResource:GBR and targetResource:NGBR) for a specific category of services as listed in Table~\ref{tab:services}. Two possible subclasses are defined in the service extension models for \gls{mc} and \gls{nmc} services requiring \gls{gbr} and \gls{ngbr} resources (e.g.,\ \textit{service:McpttGBRService} and \textit{service:NonMcpttNGBRService}).

\subsubsection{Service KPI Model}
We utilize \textit{latency, packet error rate, priority}, and \textit{5G QoS identifier (5GQI)}  as service \gls{kpi} parameters included in the \textit{icm:PropertyParameter} subclass of the \gls{icm} \gls{rdf} model. These parameters contain the values as \gls{rdf} literal terms for different services requiring the \gls{gbr} and \gls{ngbr} resources in subclasses \textit{kpi:latency}, \textit{kpi:packeterrorrate}, \textit{met:priority}, and \textit{met:qi5G}, respectively. The \gls{kpi} extension models for the \gls{mc} and \gls{nmc} services, shown in \figurename~\ref{image: tmf-icm-mcptt-nmcptt}, are based on the \glspl{kpi} from the \gls{3gpp} Table-5.7.4-1 \cite{3gpp-5gsystem-23-501}.
\import{./Resources/Tables}{services}

The values in the extension models can be extracted using SPARQL \cite{sparql-w3c} queries during the processing and translation of intents. Moreover, the information regarding expected resource type is also modeled in the service models in order to enable the orchestration of relevant resources from the available network resources. A generic framework for intent processing using the proposed knowledge base is presented in Section \ref{sec:framework}.

\subsection{Querying the Knowledge Base}
SPARQL is one of the three fundamental enablers alongside \gls{rdf} and \gls{owl} designed for querying graph-based data. In SPARQL, the queries are focused on what the user wants to know about the data rather than on the structure of the data. The SPARQL syntax utilizes specific keywords to manipulate RDF graphs and information stored in the knowledge base. For example, The keyword \textit{SELECT} matches the required data values from the RDF graphs, \textit{WHERE} describes information (relations and properties) about the queried data, and \textit{FILTER} modifies the data query. 
\begin{lstlisting}[language=SPARQL, frame=top, frame=bottom, captionpos=b, caption={Sample query template for extraction of service KPIs from the knowledge base where \textbf{param} refers to the service parameter and \textbf{serv} refers to the service being queried.}, label=lst:serv-kpi-query,
]
PREFIX icm: <http://tio.models.tmforum.org/tio/ v2.0.0/IntentCommonModel/> 

SELECT ?parameter ?value 
WHERE {
  ?service ?property [ icm:valueBy [ 
  ?parameter ?value ] ] .
  FILTER (?parameter = |\textbf{param}| && ?service = |\textbf{serv}|)
  }
\end{lstlisting}
\vspace{-0.1cm}

These common features between RDF graphs and SPARQL query structure motivate us to use SPARQL syntax to perform queries for the extraction of relevant information from the knowledge base. This is done successively for the extraction of intent templates, services, and resource information from the knowledge base, as well as the reported intent compliance information. 
A sample of the SPARQL query to retrieve service specific \glspl{kpi} from the service extension model for a generic service \textbf{serv} with associated \glspl{kpi} \textbf{param} is shown in Listing ~\ref{lst:serv-kpi-query}. 


\subsection{Applicability of the Proposed Extension Models}
A major advantage of the proposed domain-independent knowledge base is the feasibility to initialize new knowledge graphs to represent different concepts for the \gls{ibn} management and service orchestration. Thus, the proposed service and \gls{kpi} models can be extended for any type of service utilizing the \gls{icm} and having well-defined performance metrics as well as resource requirements. 
The process for expanding the proposed intent model with new service extensions is intuitive and described as:
 \begin{enumerate}
     \item The resource type is specified by defining new subclasses of \textit{icm:Target}. For instance, a new subclass could be delay critical \gls{gbr} from the \gls{5g} system specifications \cite{3gpp-5gsystem-23-501}.
     \item The service \glspl{kpi} is specified by defining the subclass of \textit{icm:PropertyParameter} for each new \gls{kpi} along with respective definitions for the \textit{icm: Expectation} class. 
     \item A new service could then be defined using the new resource target subclass of the \textit{icm:Target} and the service \glspl{kpi}. This step consists of the recommended values for the literal bounds for each \gls{kpi} parameter for different service types.
     \item The initialization of new information sources in the knowledge base is completed with the introduction of newly added parameters and service expectations in the reporting subclasses of the \gls{icm} namely \textit{icm:RequirementReporter}.
 \end{enumerate}

The described extensibility is matched by SPARQL, which provides an efficient interface to the knowledge base for retrieval of information since one can design queries that remain valid even when new services or resources are added to the network. It is also intuitive to utilize SPARQL queries to introduce performance metrics of different deployed services in the form of \textit{icm:IntentReport} in order to make them accessible in the knowledge base.

\begin{figure*}[!htbp]
\begin{center}
    \includegraphics[width=1\textwidth]{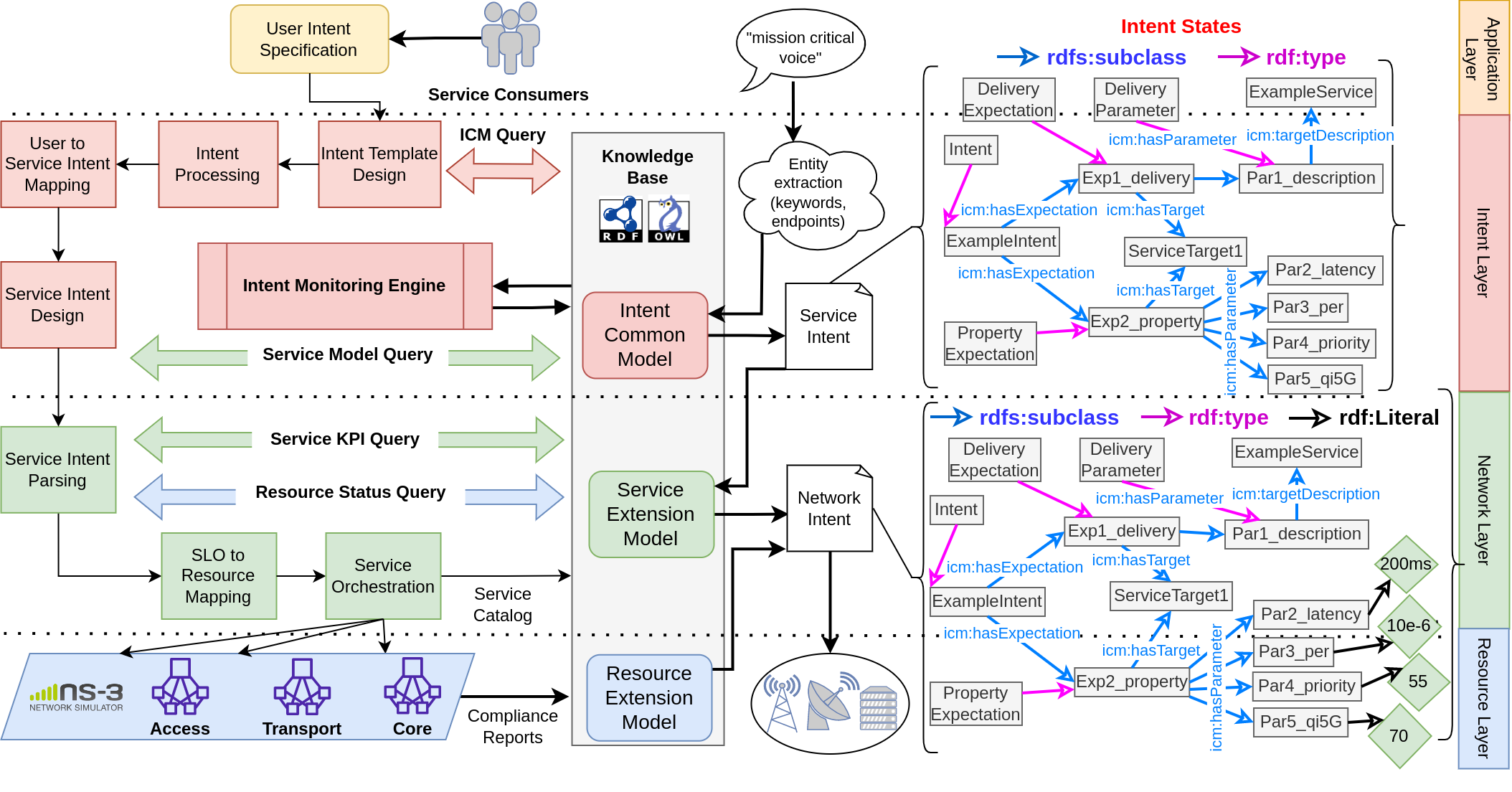}
\end{center}
\caption{Proposed intent processing framework with knowledge-based \gls{ibn} model.}
\myfigureshrinker
\label{image: ibn-framework}
\end{figure*}

%% file: Resources/Tables/services.tex
\begin{table}
\caption{Expected service KPIs \cite{3gpp-5gsystem-23-501}.}
\label{tab:services}
\begin{adjustbox}{width=\columnwidth,center}
\begin{tabular}{lcccccc}
\midrule
\multicolumn{1}{c}{\textbf{\begin{tabular}[c]{@{}c@{}}Service\\ Example\end{tabular}}} &
  \textbf{\begin{tabular}[c]{@{}c@{}}5GQI \\ Value\end{tabular}} &
  \textbf{\begin{tabular}[c]{@{}c@{}}Resource\\ Type\end{tabular}} &
  \textbf{\begin{tabular}[c]{@{}c@{}}Priority\\ Level\end{tabular}} &
  \textbf{\begin{tabular}[c]{@{}c@{}}Latency\\ (ms)\end{tabular}} &
  \textbf{\begin{tabular}[c]{@{}c@{}}Packet Error \\ Rate (PER)\end{tabular}} &
  \textbf{\begin{tabular}[c]{@{}c@{}}Service \\ Category\end{tabular}} \\ \midrule
Conversational Voice & 1  & GBR  & 20 & 100 & 10e-2 & NMC \\ \midrule
Conversational Video & 2  & GBR  & 40 & 150 & 10e-3 & NMC \\ \midrule
Process Monitoring   & 3  & GBR  & 30 & 50  & 10e-3 & NMC \\ \midrule
IMS Signaling        & 5  & NGBR & 10 & 100 & 10e-6 & NMC \\ \midrule
Buffered Video       & 6  & NGBR & 60 & 300 & 10e-6 & NMC \\ \midrule
Live Video           & 7  & NGBR & 70 & 100 & 10e-3 & NMC \\ \midrule
MCPTT Up Voice       & 65 & GBR  & 7  & 75  & 10e-2 & MC  \\ \midrule
MCPTT Up Video       & 67 & GBR  & 15 & 100 & 10e-3 & MC  \\ \midrule
NonMCPTT Voice          & 66 & GBR  & 20 & 100 & 10e-2 & MC  \\ \midrule
MCPTT Data           & 70 & NGBR & 55 & 200 & 10e-6 & MC  \\ \midrule
MCPTT Signaling      & 69 & NGBR & 5  & 60  & 10e-6 & MC  \\ \midrule
\end{tabular}
\end{adjustbox}
\vspace{-0.2cm}
\end{table}

%% file: Sections/3-Sec3.tex
\section{An Intent Processing Framework with Contextual Knowledge Base} \label{sec:framework}
In this section, we propose a knowledge-based \gls{ibn} framework for operational management of multi-domain next-generation networks offering heterogeneous services as shown in \figurename~\ref{image: ibn-framework}. The proposed extension facilitates an organized intent processing lifecycle and knowledge management framework that will help in translating high-level intent templates from the service consumers into specific technical artifacts to be used by the network administrator and orchestrator.

\subsection{Application Layer}
The expression and definition of intents for the orchestration of different services and network functions are achieved through an application portal with a catalog of offered services to consumers. In the proposed setup, the service consumers specify the type of expected service using pre-defined options such as `Mission Critical Voice' at the application portal. Therefore the purpose of the application layer is to provide the user an option to express the intent service requirements.

The application layer also contributes to the knowledge base with the available service profiles of active service consumers.
\subsection{Intent Layer}
The intent layer is responsible for providing the required context to aid the modeling and translation of consumer intents. The input intent by a user application is analyzed to create an \gls{rdf} model of intent by utilizing the \gls{tmf}'s \gls{icm}. This model consists of the bare minimum components of intent for operational, \textbf{monitoring}, and \textbf{reporting} purposes. The operational part consists of the expected \glspl{slo} and the acceptable range of performance metric values. The monitoring part is concerned with the fulfillment of intents by the network functions. The reporting part defines the set of \glspl{slo} to consider while monitoring the lifecycle of an intent after deployment.

A user's intent is refined by adding required details during the \textbf{design of the service intent} via queries to the intent model to extract information (target resources, \glspl{slo}, etc) relevant to the requested services. Starting from the \gls{icm} intent template, extensions are added to create an empty service intent with the required expectations for the expected service. 

An example of a service intent modeled from a service requested by a user is depicted in \figurename~\ref{image: ibn-framework}. It includes the proposed service and resource extension models with knowledge facts describing the expectations of the service in terms of \textit{icm:DeliveryExpectation} and \textit{icm:PropertyExpectation}, corresponding to fields from the \gls{icm}.  After verification, the requested services and their respective parameters are embedded in the \gls{icm} to complete the service intent.
The intent monitoring and reporting is modeled through the subclasses of the \textit{icm:RequirementReporter} realized as \textit{icm:IntentReport} and \textit{icm:ExpectationReport} in  the \gls{icm}. The \glspl{slo} for a specific service is expressed through the extension models of \textit{icm:Expectation} and \textit{icm:Parameter} classes. The report consists of several events representing the state of the intent throughout the intent management lifecycle. Key intent state events are \textit{icm:IntentStateReceived, icm:IntentStateCompliant, icm:StateDegraded, icm: StateUpdated} and \textit{icm:StateFinalized}. 


\begin{figure*}[!htbp]
     \centering
     \begin{subfigure}{0.49\textwidth}
         \centering
         \includegraphics[width=\textwidth]{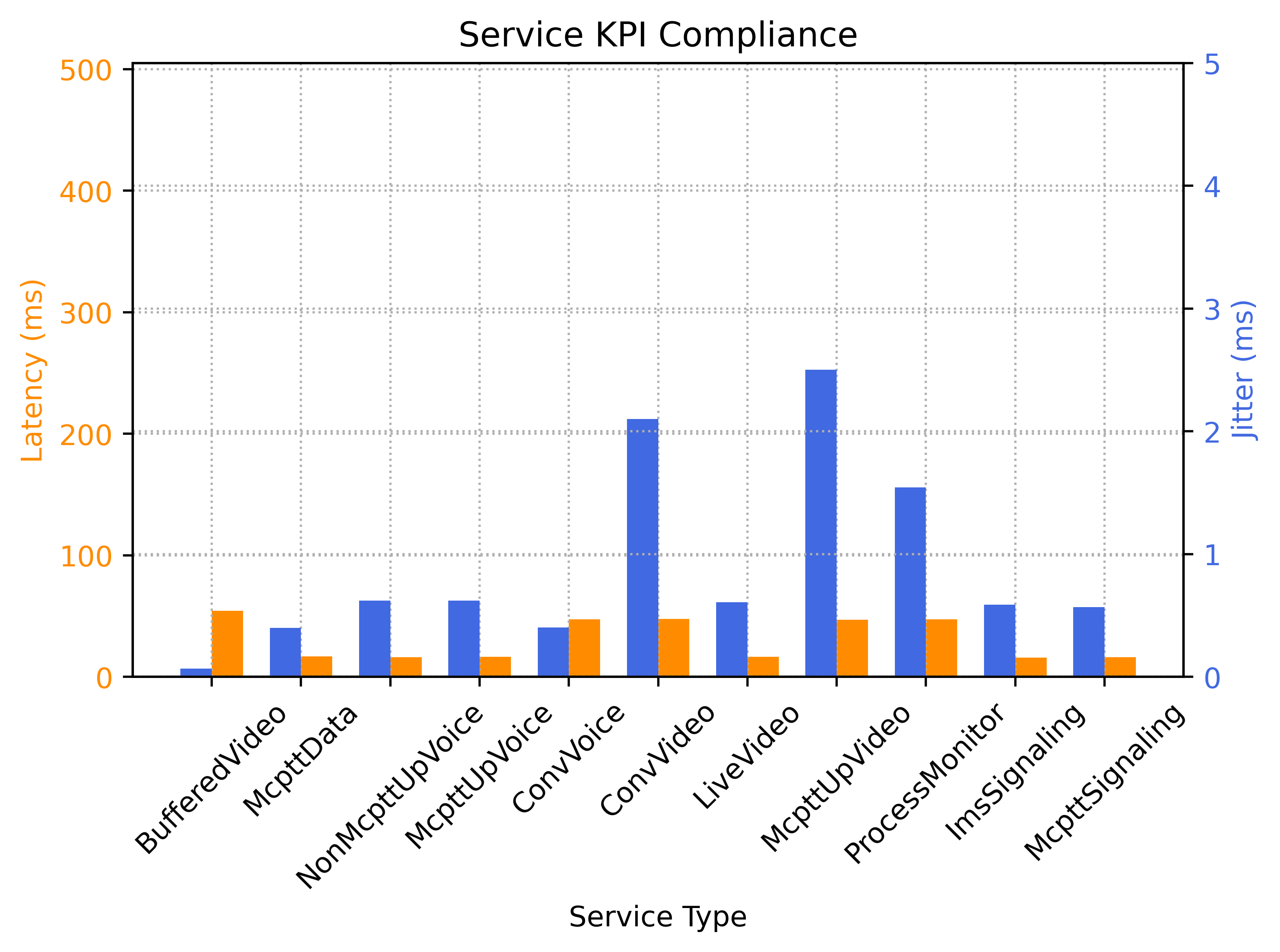}
         \caption{Non-congested}
         
         \label{image: service-timing}
     \end{subfigure}
     \hfill
     \begin{subfigure}{0.49\textwidth}
         \centering
         \includegraphics[width=\textwidth]{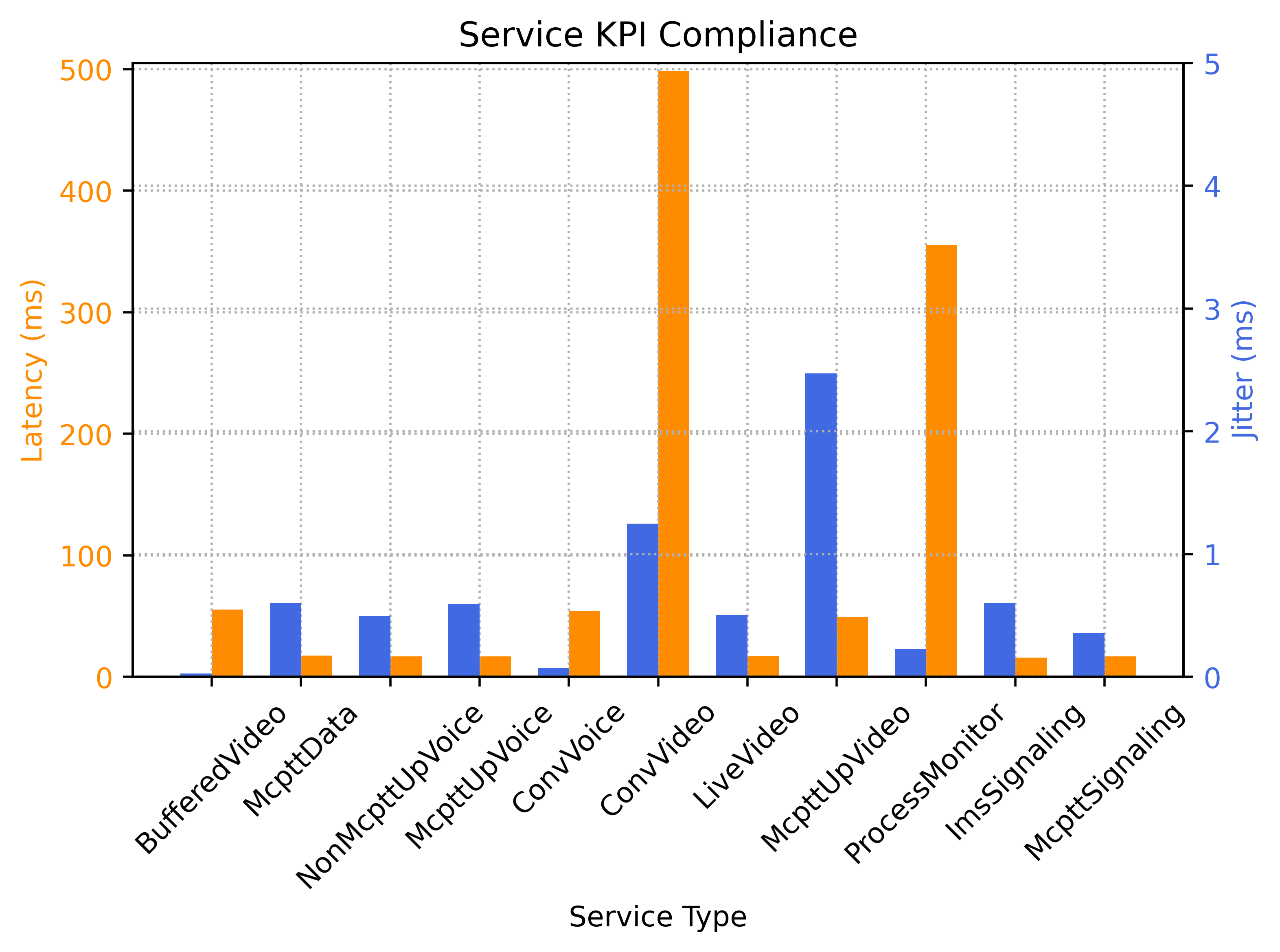}
         \caption{Congested}   
         \label{image: service-cong-timing}
     \end{subfigure}
     \caption{KPI performance for deployed services.}
         \myfigureshrinker
\end{figure*}
\subsection{Network Layer}
 The network layer consists of processing of service intents and converting them into network intents readily deployed using a service orchestrator. The service intent is forwarded to the network layer for validation and orchestration as per available resources and service models. This is accomplished using a network created by querying the service extension model from the knowledge base for \glspl{slo}. A sample of the network intent is shown in \figurename~\ref{image: ibn-framework} consisting of the required information for a service to be deployed.
The validation at the network layer detects possible conflicts with respect to resource allocation to the requested service \glspl{kpi} given the state of the available resources accomplished using the resource validation query. 

\subsection{Resource Layer}
The resource layer consists of the underlying infrastructure and associated resources for the deployment of different services requested by the network intents via the service orchestrator. The resource model information is included in the knowledge base in the form of resource type, as an extension and subclass of the \textit{icm:Target} class shown in \figurename~\ref{image: tmf-icm-mcptt-nmcptt}. The proposed extension models for the resource model in the \gls{icm} consist of two types of \lq icm:targetResources\rq, namely NGBR and GBR. 

In addition, the resource layer is also responsible for keeping a catalog for the available and utilized resources, as well as the compliance of the deployed services according to the \glspl{slo}. This information is updated in the knowledge base and made available upon request for intent monitoring and reporting. 

%% file: Sections/4-Sec4.tex
\section{Performance Evaluation via Intent-Driven \gls{5g} Service Orchestration} \label{sec: results}
This section covers an \gls{ibn} proof-of-concept using \gls{ns-3} to orchestrate \gls{5g} services in a non-standalone network. The simulated setup follows the workflow at the different layers shown in \figurename~\ref{image: ibn-framework}. The knowledge base and intent processing framework are implemented in Python, and the proposed service models and associated queries are made publicly available on github \cite{intentRDF23}. 
The intents are deployed using a \gls{ns-3} network consisting of \gls{5g} new radio (NR) access network and \gls{lte} core. 

\subsection{Simulation Setup}
The simulation setup consists of several steps following the proposed \gls{ibn} framework from \figurename~\ref{image: ibn-framework} summarized as:
\begin{enumerate}
    \item Acquisition of intent from the user (\textit{application layer});
    \item Recognition of relevant keywords in the user intent and \gls{icm} query from the knowledge base (\textit{intent layer});
    \item Querying the service extension model and creation of a service intent (\textit{intent layer});
    \item Service \gls{kpi} query from the service extension model and creation of the network intent (\textit{network layer});
    \item Deployment of the network intent in the underlying \gls{ns-3} network (\textit{resource layer}).
    \item Intent status reporting with compliance information of deployed services to the intent-generating user and manager.
\end{enumerate}


\subsection{Simulation Results}
The services are orchestrated in two different scenarios; (a) normal operational conditions with specified \glspl{kpi} and (b) a congested network with accumulated data queues for different services. We have analyzed the proposed \gls{ibn} processing framework by observing the creation of intents and the associated interactions with the knowledge base. We evaluate the compliance of the deployed intent-based services with the \glspl{kpi} from the \gls{3gpp} Table-5.7.4-1 \cite{3gpp-5gsystem-23-501} depicted in Table I.

\subsubsection{Service Compliance to Intents}
The proposed \gls{ibn} framework for intent processing and deployment requires recursive testing for different scenarios and types of services. 
The network intents for the services in Table I are generated and deployed, and the performance measurements presented in \figurename~\ref{image: service-timing} show that the \gls{pdb} or experienced latency remain within an acceptable range with minimum jitter. 
The reporting parameters continue to periodically update the intent monitoring module of the status of the services to confirm compliance with the network intents. This event is a subclass of \textit{icm:event}, and the \gls{imo}\cite{tmforum-intent-mgmt-ont-tr292} defines it as \textit{imo:StateComplies} for the compliant services.  

\subsubsection{Non-Compliance of Deployed Services}
We have also deployed the intents for different services in a congested network deployment in order to visualize the feedback mechanism towards the intent monitoring engine. The results obtained for the deployed intents in this scenario are shown in \figurename~\ref{image: service-cong-timing}.

The congestion is obtained by increasing the packet sizes for the services in order to produce a backlog of packets in queues. We observe that 9 out of 11 deployed services via intents remain compliant where the services \textit{icm:ConvVideo} and \textit{icm:ProcessMonitor} violate their respective \lq latency\rq \glspl{slo}. The expected latency threshold for \textit{ConvVideo} and \textit{ProcessMonitor} services are 150 and 50 milliseconds respectively. Furthermore, we furnish measured values of the observed jitter for the implemented services to facilitate the visualization of a comprehensive latency profile. We compare the compliance of deployed services with their respective network intents by analyzing the generated intent reports.
\begin{lstlisting}[language=SPARQL, frame=top, frame=bottom, captionpos=b, caption={RDF snippet from the \textit{Intent Report} generated for the \textbf{ConvVideo} service.}, label=lst:iReport-convVideo,
]
rep:ER2_ServiceProperty a icm:ExpectationReport ;
  |\color{blue}\textbf{icm:compliant}| [ a icm:PropertyParameter ;
    icm:reason |\textbf{icm:ReasonMeetsRequirement}| ;
    icm:reportsAbout exI:Par3_per ;
    icm:valueBy [ |\color{blue}\textbf{kpi:packeterrorrate}| "0"^^xsd:string ] ] ;
  |\color{red}\textbf{icm:degraded}| [ a icm:PropertyParameter ;
    icm:reason |\textbf{icm:ReasonNotCompliant}| ;
    icm:reportsAbout exI:Par2_latency ;
    icm:valueBy [ |\color{red}\textbf{kpi: latency}| "493.1097 ms"^^xsd:string ] ] ;
  icm:hasTarget catalog:ExampleService ;
  icm:reportsAbout exI:Exp1_property .
\end{lstlisting}

\begin{lstlisting}[language=SPARQL, frame=top, frame=bottom, captionpos=b, caption={RDF snippet from the \textit{Intent Report} generated for the \textbf{McpttData} service.}, label=lst:iReport-mcpttData
]
rep:ER2_ServiceProperty a icm:ExpectationReport ;
  |\color{blue}\textbf{icm:compliant}| [ a icm:PropertyParameter ;
    icm:reason |\textbf{icm:ReasonMeetsRequirement}| ;
      icm:reportsAbout exI:Par3_per ;
      icm:valueBy [ |\color{blue}\textbf{kpi:packeterrorrate}| "0"^^xsd:string ] ],
    [ a icm:PropertyParameter ;
      icm:reason |\textbf{icm:ReasonMeetsRequirement}| ;
      icm:reportsAbout exI:Par2_latency ;
      icm:valueBy [ |\color{blue}\textbf{kpi:latency}| "17.6459 ms"^^xsd:string ] ] ;
  icm:hasTarget catalog:ExampleService ;
  icm:reportsAbout exI:Exp1_property .
\end{lstlisting}
\subsubsection{Intent Status and Reporting}
Intent reporting forms the key aspect of the intent management lifecycle and it is implemented using the \textit{icm:IntentReport} class from the \gls{icm}. The service parameters from the deployed intent are utilized to populate the \textit{icm:RequirementReporter} class using the \textit{icm:ReportingParameter} subclass of the \textit{icm:Parameter}. The \gls{ns-3} simulation is monitored for the reporting parameters with the observed values being added to the generated \textit{icm:IntentReport} for different deployed services. The compliance of the service to a particular parameter is specified through \textit{icm: compliant} and \textit{icm:degraded} properties. The reason for the evaluation of a particular service \gls{kpi} (icm:Parameter) is specified as \textit{icm:ReasonNotCompliant} for degraded and \textit{icm:ReasonMeetsRequirement} for complying values respectively. We provide an RDF representation from the intent reports for the \textbf{ConvVideo} and \textbf{McpttData} services in Listings~\ref{lst:iReport-convVideo} and~\ref{lst:iReport-mcpttData}, respectively.

The \textit{kpi:latency} and \textit{kpi:packeterrorrate} values are reported in the intent report for the \textit{ConvVideo} service as compliant or degraded to the knowledge base. The \gls{imo} defines that an \textit{imo:StateDegrades} event notifies the intent monitoring engine of the performance degradation for the services failing to meet its \glspl{kpi}.

Similarly, the intent report for the \textit{McpttData} service shows that both of the observed \glspl{kpi} are compliant and the intent state is reported as \textit{imo:StateComplies} to the knowledge base and intent monitoring module. An excerpt from the intent report is provided in Listing ~\ref{lst:iReport-mcpttData}.

%% file: Sections/5-Sec5.tex
\section{Conclusion} \label{sec: conc}
This paper covers intent representation and knowledge organization in \gls{ibn} using a service orchestration use case for cellular \gls{5g} networks and a common knowledge base for the service, intent, and resource models is proposed. We focus on the intent representation and processing in the networking domain assuming the availability of service information mapped from the declarative user intent expression. The proposed \gls{ibn} framework provides the ability to translate intents into various network comprehensible forms leading to the deployment of the requested services. Different network intents are deployed as distinct services using \gls{ns-3} based infrastructure and the observed \glspl{kpi} are compliant with the expressed \glspl{slo} of services. 

\vspace{0.5cm}
